\documentclass[letterpaper, 10 pt, conference, compsocconf]{IEEEtran}
\IEEEoverridecommandlockouts
\usepackage{cite}
\usepackage{amsmath,amssymb,amsfonts}
\usepackage{algorithmic}
\usepackage{graphicx}
\usepackage{textcomp}
\usepackage{xcolor}
\def\BibTeX{{\rm B\kern-.05em{\sc i\kern-.025em b}\kern-.08em
    T\kern-.1667em\lower.7ex\hbox{E}\kern-.125emX}}
\begin{document}

\IEEEoverridecommandlockouts
\IEEEpubid{\makebox[\columnwidth]{ 978-1-7281-7172-2/20\$31.00 \copyright 2020 IEEE \hfill} \hspace{\columnsep}\makebox[\columnwidth]{ }}
\title{Undergraduate Student Research With Low Faculty Cost
}

\author{\IEEEauthorblockN{Sindhu Kutty}
\IEEEauthorblockA{\textit{Division of Computer Science and Engineering} \\
\textit{University of Michigan}\\
Ann Arbor, USA \\
skutty@umich.edu}
\and
\IEEEauthorblockN{Mark Guzdial}
\IEEEauthorblockA{\textit{Division of Computer Science and Engineering} \\
\textit{University of Michigan}\\
Ann Arbor, USA \\
mjguz@umich.edu}
}

\maketitle
\begin{abstract}
Undergraduates are unlikely to even consider graduate research in Computer Science if they do not know what Computer Science research is. Many programs aimed at introducing undergraduate to research are structured like graduate research programs, with a small number of undergraduates working with a faculty advisor. Further, females, under-represented minorities, and first generation students may be too intimidated or the idea of research may be too amorphous, so that they miss out on these programs. As a consequence, we lose out on opportunities for greater diversity in CS research. 

We have started a pilot program in our department where a larger number of students (close to two dozen) work with a single faculty member as part of a research group focused on Machine Learning and related areas. The goal of this program is not to convince students to pursue a research career but rather to enable them to make a more informed decision about what role they would like research to play in their future.

In order to evaluate our approach, we elicited student experience via two anonymized exit surveys. Students report that they develop a better understanding of what research in Computer Science is. Their interest in research was increased as was their reported confidence in their ability to do research, although not all students wanted to further pursue computer science research opportunities. Given the reported experience of female students, this program can offer a starting point for greater diversity in CS research.
\end{abstract}


\section{Introduction}
Students who have only a second-hand or vague understanding of the Computer Science research process may never seriously consider research as a possible career path. This can hinder engagement from students who are in demographic groups that are under-represented in the field, including women, racial minorities and first generation students.

In this project, we are interested in providing a safe space for students to `try research on for size'. We aim to provide enough structure to the process so that it is scalable. We hypothesize that we can use this model to target students who may not have had exposure to the idea of research or might have been intimidated by the idea.

\section{Project Description}
This project is an educational initiative to introduce undergraduates to research in an environment that is both more scalable and less intimidating than common models of undergraduate research opportunities. Lack of awareness or a self-perceived lack of preparedness may mean that some students miss out on these programs \cite{b2}. This program deconstructs the research process via twice-weekly meetings, a guided literature survey and a replication project. Unlike UROP and similar programs, the faculty to student ratio is low (one faculty to 22 students in this offering). Given concerns of high enrollment, this program provides a scalable approach to introduce undergraduates to CS research. 

In order to evaluate the effectiveness of our approach, we elicited student experience via two anonymized exit surveys. Students report having a better sense of what is involved in doing CS research despite not having a clear idea previously. They found the level of detail involved in the process challenging but enjoyable. 

\section{Elements of the Program}
This program was advertised to students towards the end of a high enrollment upper-level undergraduate course on Machine Learning taught by one of the authors. The course was run for a duration of 7 weeks with twice-weekly meetings for the first six weeks and a final presentation in the last week. Optionally, students were encouraged to submit a written report of their work. This was an informal course offering: in particular, students received no academic credit or grade for participation in this course. 

Elements of this methodology have been used previously in other institutions to engage students with research.  Replication projects have been used to teach research in STEM fields, including mathematics, psychology and even computer science (see, for instance, \cite{b1}). Reading groups and seminar-style courses typically have assigned readings and presentations. However, we have provided a framework that uses these elements to introduce students to CS research with low per-student faculty cost.

The course focused on recent research results in Machine Learning and Social Computing. A sampling of subtopics were presented in the kick-off meeting. Students then divided into self-selected groups of about 3 each based on topic preferences. If larger groups formed around topics, students were divided into smaller subgroups. 

\subsection{Literature Survey Component}
To fully immerse students in the field, students do a literature survey of different subareas. The faculty mentor works with each group to pick representative papers for each subtopic, by pointing out some sample results and relevant and reputable conferences. The group that selected the subtopic was responsible for presenting the main results in these papers to the entire cohort. All groups read papers from every subtopic. A template is provided to students to guide them in reading and analyzing the paper, and all groups produce a written summary of papers that  they have read that week. 

\subsection{Replication Project Component}
To engage students in thinking about research contributions, we have students design a replication study.
The faculty mentor  works with each group to pick an appropriate (set of) result(s) for replication. For instance, students may choose a highly theoretical result and design an experiment to understand the result and test the impact of relaxing the assumptions in the results. The faculty mentor meets with each small group on a weekly basis to discuss their progress and provide feedback on their work. In order to approach this systematically, the tasks are structured so that each week's assignment is of increasing specificity. For instance, one of the first steps may be to delineate the results for replication. In a later week, the task may involve specifying pseudocode to test their hypotheses.

\section{Highlights of Survey Results}
There were 22 members of the research group. Two anonymized exit surveys were administered approximately a month and 3 months after the final presentation. We received 19 responses to the first survey and 10 to the second. 3 out of 10 respondents in the second survey were female. As a point of comparison, the general population of female CS undergraduate students at our institution is ~22\%. We do not have gender data for the first survey.

We elicited students' interest and confidence in doing research via a 5-point Likert scale. Students reported an increased interest and confidence in research. Most students had some exposure to reading research papers but their understanding of what the research process entailed in Computer Science was vague: ``\textit{I thought it is involved a lot coding at least and a lot about implementation in real life.}". They were surprised the amount of detail and depth involved in the process: ``\textit{The research process focuses a lot more on the thought experiment and ideas behind. The ability of asking interesting question is crucial in research.}''. They noted that they had a better understanding of the breadth of applications, likely as a consequence of exposure to interdisciplinary work. They also report having formed a better understanding of the research process: ``\textit{Yes, I have a much better idea. Especially after working on a replication project I got a reasonable idea of what CS research looks like.}" and  indicate that the course can help other students evaluate whether CS research is right for them: ``\textit{Yes I strongly agree that it is helpful regardless of students research status. In previous courses, I feel majority of the time was spent on how to implement the code and following the task instructions. But this reading group really engages me to think about the reasons and broader picture behind it. I enjoy this challenging process.}" 

In particular, 2 out of 3 female respondents reported that they definitely saw CS research as part of their future plans. All female respondents also reported having a better sense of what CS research is about due to the program. They reported that the program had helped them decide their plans.  All female respondents also report that a student who is not currently interested in research might benefit from such a program: ``\textit{If they haven't had that kind of exposure being exposed to research papers and doing these sorts of projects may turn them around if they discover that they enjoy it}"

\section{Conclusions and Future Work}
After engaging in this program students report that they developed a better understanding of what research in Computer Science is. They reported enjoying the course, indicating a more welcoming  environment. Their reported confidence in their ability to do research was increased. Overall, participants also reported having an increased interest in research. All respondents reported that, to varying degrees, it helped inform their post graduation plans. Not all female respondents reported that they necessarily saw CS research as part of their future plans; however, they all reported that the program had helped them decide their plans. All female respondents also reported having a better sense of what CS research is about due to the program and would recommend it to a student who is unsure about CS research. Given the reported experience of female students, this program can offer a starting point for greater diversity in CS research. 
Ours is an R1 institute with high enrollments in CS. 
We would like to evaluate the suitability of this approach to different departments with potentially different demographics and cultures. 
It would also be interesting to explore whether this program can be effectively adapted to other areas of CS than Machine Learning and related subfields.

\section*{Acknowledgments}

The authors would like to thank Kathryn Cunningham for help with processes related to IRB and the students who participated in the course and provided feedback. The first author gratefully acknowledges the 2019 Course Development Fund, CSE department, University of Michigan.


\vspace{12pt}



\end{document}